\documentclass[twoside]{article}
\usepackage[cp1251]{inputenc}
\usepackage[english,russian]{babel}
\usepackage{fam}
\usepackage{makeidx}
\usepackage{ifthen}
\usepackage{array, hhline}
\usepackage{graphicx}
\usepackage{graphpap}
\usepackage{amsfonts}
\usepackage{amsmath}
\usepackage{amscd}
\usepackage{amssymb}
\usepackage{amstext}
\usepackage[mathscr]{eucal}
\usepackage{latexsym}
\usepackage[centerlast,footnotesize]{caption2}
\usepackage{texdraw}
\usepackage{xcolor}
\usepackage{color}
\usepackage{url}
\usepackage[breaklinks]{hyperref}
\usepackage{hyperref}
\hypersetup{
    colorlinks,
    citecolor=black,
    filecolor=black,
    linkcolor=black,
    urlcolor=teal
}

\begin{document}
\normalsize 
\baselineskip=1.2\baselineskip

\setcounter{page}{51}
\clearpage
\thispagestyle{empty}

\myTitle{Eventological $H$-theorem}

\colontitleeng{Vorobyev}

\myAuthorAddress{Oleg Yu. Vorobyev}{Siberian Federal Univdersity\\
Institute of Mathematics and Computer Science\\
oleg.yu.vorobyev@gmail.com\\
ovorobov@sfu-kras.ru\\
http://www.academia.edu/OlegVorobyev}

\footavteng{Oleg Yu. Vorobyev}

\renewcommand{\thefootnote}{\arabic{footnote}}
\setcounter{footnote}{0} 
\setcounter{equation}{0}
\setcounter{figure}{0} 
\setcounter{table}{0}
\setcounter{section}{0} 
\setcounter{defini}{0}
\setcounter{thm}{0}

\begin{abstracten}
\begin{center}
\begin{minipage}{9.8cm}
\footnotesize
We prove the eventological $H$-theorem that complements the Boltzmann $H$-theorem from statistical mechanics \cite{6-Boltzmann1872}
and serves as a mathematical excuse (mathematically no less convincing than the Boltzmann $H$-theorem for the second law of thermodynamics)
for what can be called ``the second law of eventology'', which justifies the application of \emph{Gibbs and ``anti-Gibbs'' distributions} \cite{6-Vorob'ov2007} of sets of events minimizing relative entropy, as statistical models of the behavior of a rational subject, striving for an equilibrium eventological choice between perception and activity in various spheres of her/his co-being.
\end{minipage}
\end{center}
\end{abstracten}

\section{Eventological $H$-theorem\protect\\ on extreme properties of Gibbs and ``anti-Gibbs''\protect\\ eventological distributions}

{\bf Theorem (eventological $H$-theorem)}. {\it Let $(\Omega,
\mathscr{F}, \mathbf{P})$ be the eventological space, $\frak{X} \subseteq
\mathscr{F}$ be the finite set of events, ${\mathscr V}(X)$ be nonnegative bounded
function\footnote{In eventology, the ${\mathscr V}(X)$ is interpreted as a value (for a rational subject) of the set of events
 $X \subseteq \mathfrak{X}$ occurrence on $(\Omega, \mathscr{F}, \mathbf{P})$.} on $2^\frak{X}$,
$p_*(X)$ be some fixed eventological distribution on $2^\frak{X}$, and let
eventological distributions $p(X)$ on $2^\frak{X}$ keep a mean value of the function ${\mathscr V}(X)$ at the given level
$$
\langle {\mathscr V} \rangle = \sum_{X \subseteq \frak{X}} p(X)
{\mathscr V}(X). \eqno{(\mathscr{V})}
$$
Then the minimum of relative entropy
$$
H_{\frac{p}{p_*}} = \sum_{X \subseteq \frak{X}} p(X) \ln
\frac{p(X)}{p_*(X)} \to \min_p
$$
among all eventological distributions $p$  is achieved on Gibbs and anti-Gibbs eventological distributions of the following form:
$$
p(X) = \frac{1}{Z_{p_*}} \exp \Big\{ - \beta {\mathscr V}(X)\Big\}
p_*(X), \ \ \ X \subseteq \frak{X}, \ \ \ \beta \geq 0,
$$
$$
p(X) = \frac{1}{Z_{p_*}} \exp \Big\{ \gamma {\mathscr V}(X)\Big\}
p_*(X), \ \ \ X \subseteq \frak{X}, \ \ \ \gamma \geq 0,
$$
which can be rewritten without a normalizing factor $1/Z_{p_*}$ in
the equivalent form:
$$
\frac{p(X)}{p(\emptyset)} = \exp \Big\{ - \beta ({\mathscr
V}(X)-{\mathscr V}(\emptyset))\Big\} \frac{p_*(X)}{p_*(\emptyset)}, \ \
\ X \subseteq \frak{X},
$$
$$
\frac{p(X)}{p(\emptyset)} = \exp \Big\{ \gamma ({\mathscr
V}(X)-{\mathscr V}(\emptyset))\Big\} \frac{p_*(X)}{p_*(\emptyset)}, \ \
\ X \subseteq \frak{X}.
$$
}

{\bf P r o o f} \ uses the idea of proof of one variant of Boltzmann $H$-theorem from statistical mechanics (in the formulation taken from \cite[p. 41]{6-Isihara1971}). As it turned out, this long-standing idea is enough to get much more general conclusions under classical assumptions.

Let us compare the relative entropy for the Gibbs factor
$$
f(X) = \exp \Bigg\{ -\beta \mathscr{V}(X) \Bigg\} p_*(X), \ \ \ \beta
\geq 0,
$$
or for the \emph{Gibbs ``anti-factor''}\footnote{The term is proposed by me,
has no analogs in statistical physics.}
$$
f(X) = \exp \Bigg\{ \gamma \mathscr{V}(X) \Bigg\} p_*(X), \ \ \ \gamma
\geq 0,
$$
by introducing the general notation for them\footnote{In this case, we always have $p(X) =
\frac{1}{Z_{p_*}} f(X)$.}
$$
f(X) = \exp \Bigg\{ \alpha \mathscr{V}(X) \Bigg\} p_*(X)
$$
(where $\alpha \in \mathbf{R}$ is a real arbitrary-sign parameter), with the relative entropy for any function
$\varphi(X)$ that is normalized to the same factor $Z_{p_*}$ as the function $ f (X) $ is normalized to.

Introducing a new function $g(X)$ such that $\varphi(X) = f(X) \cdot g(X)$, we find
$$
H_{\frac{f}{p_*}} - H_{\frac{\varphi}{p_*}} = \frac{1}{Z_{p_*}}
  \sum_{X \in 2^{\frak X}} \left[
    f(X) \ln \frac{f(X)}{p_*(X)}
  - \varphi(X) \ln \frac{\varphi(X)}{p_*(X)}
               \right] =
$$
$$
= \frac{1}{Z_{p_*}}
  \sum_{X \in 2^{\frak X}} f(X) \left[
        \ln \frac{f(X)}{p_*(X)}
        - g(X) \ln \frac{f(X)g(X)}{p_*(X)}
       \right].
\eqno{(B0)}
$$
The normalization of probabilities gives
$$
\sum_{X \in 2^{\frak X}} \Bigg[\varphi(X)-f(X)\Bigg] = \sum_{X \in 2^{\frak X}}
f(X) \Bigg[g(X)-1 \Bigg] = 0, \eqno{(B1)}
$$
and the theorem condition $(\mathscr{V})$ gives
$$
\sum_{X \in 2^{\frak X}} \Bigg[\varphi(X)-f(X)\Bigg] \mathscr{V}(X) = 0.
\eqno{(\mathscr{V}1)}
$$
Given that
$$
\alpha \mathscr{V}(X) = \ln \frac{f(X)}{p_*(X)},
$$
we obtain from $(\mathscr{V}1)$ and $(B1)$
$$
\frac{1}{Z_{p_*}} \sum_{X \in 2^{\frak X}}
 \Bigg[\varphi(X)-f(X)\Bigg] \ln \frac{f(X)}{p_*(X)} =
 $$
 $$
= \frac{1}{Z_{p_*}} \sum_{X \in 2^{\frak X}} f(X) \Bigg[g(X)-1\Bigg]
  \ln \frac{f(X)}{p_*(X)} = 0.
\eqno{(B2)}
$$
Subtracting $(B1)$ and $(B2)$ from $(B0)$, we obtain
$$
H_{\frac{f}{p_*}} - H_{\frac{\varphi}{p_*}} = - \frac{1}{Z_{p_*}}
  \sum_{X \in 2^{\frak X}} f(X) \Bigg[g(X)\ln g(X) -g(X)+1\Bigg].
$$
By the definition, the function $f(X)$ is positive, and the variable
$$
\Bigg[g \ln g - g +1\Bigg] = \int_1^g  \ln g dg
$$
is non-negative for any positive $g$. Hence, $H_{\frac{f}{p_*}} - H_{\frac{\varphi}{p_*}} \leq 0 $,
i.e. the function $H_{\frac{\varphi}{p_*}}$ is always not less than $H_{\frac{f}{p_*}}$. The theorem is proved.

\section{Interpretations of the eventological $H$-theorem}

\subsection{The direct analogy with physical interpretation}

When the relative entropy $H_{\frac{p}{p_*}}$ of a physical system (with distribution $p$) relative to the environment (with distribution
$p_*$) has a minimum value, in statistical thermodynamics it is considered that the system is in equilibrium with the surrounding medium, and its decrease with time corresponds to an approximation to equilibrium with a given medium\footnote{This is the \emph{principle of minimum relative entropy of the system for a fixed level of entropy of the environment} equivalent to the \emph{maximum entropy principle} --- the cornerstone of the second law of thermodynamics: ``the increase in the entropy of the system as it approaches to equilibrium''. ``The change of the maximum to the minimum'' is explained by formal differences in the sign between the traditional definitions of entropy: $\sum_{X \subseteq \frak{X}} p(X) \ln p(X)$ and relative entropy: $\sum_{X\subseteq\frak{X}} p(X) \ln(p(X)/p_*(X))$.}.

This physical analogy is used by us to construct a more general eventolo\-gical model of the behavior of a rational subject
based on the idea of an equilibrium choice between the \emph{perception \emph{and} activity}, to which the rational subject is doomed in the process of co-being.
When the relative entropy $H_{\frac{p}{p_*}}$ of a rational subject (with distribution $p$) relative to one's own tastes and preferences (with $p_*$ distribution) is minimal, in eventology it is considered that the rational subject is in equilibrium with her/himself, and \emph{the decrease in relative entropy with time corresponds to the striving of the rational subject to balance with her/himself --- ``the second law of eventology''}.

Therefore, the proved eventological $H$-theorem serves as a mathematical excuse\footnote{Mathematically no less convincing than the Boltzmann $H$-theorem for the second law of thermodynamics.} for what can be called ``the second law of eventology '', which justifies the use of {Gibbs and anti-Gibbs distributions} \cite{6-Vorob'ov2007} of sets of events that minimize relative entropy as statistical models of the behavior of the rational subject, striving for an equilibrium eventological choice between perception and activity in various spheres of her/him co-being.

\subsection{Event interpretation}

In the eventological $H$-theorem, as in its physical predecessor,
it is considered we know the following distribution
$$
\mathscr{V}(X), \ X \subseteq \mathfrak{X},
$$
of the set-function of value $\mathscr{V}$ of subsets of events from $\mathfrak{X}$ that is defined on $2^\mathfrak{X}$,
and the fixed eventological distribution
$$
p^*(X), \  X \subseteq \mathfrak{X}
$$
of the set of events $\mathfrak{X}$.
Then the range of considered eventological distributions of the set of events $\mathfrak{X}$ is limited by such eventological distributions
$$
\{p(X), \  X \subseteq \mathfrak{X}\},
$$
that satisfies the restriction
$$
\langle \mathscr{V} \rangle = \sum_{X \subseteq \mathfrak{X}}p(X)
\mathscr{V}(X).
$$
In other words, these eventological distributions ``keep'' the $p$-mean value of set-function of value $\mathscr{V}$ at a fixed level $\langle \mathscr{V} \rangle$:
$$
\mathbf{E}_p (\mathscr{V}) = \sum_{X \subseteq \mathfrak{X}}p(X)
\mathscr{V}(X) = \langle \mathscr{V} \rangle.
$$
The theorem states that among the given eventological distributions the relative entropy $\mathcal{H}_{\frac{p}{p^*}}$
reaches a minimum on Gibbs and anti-Gibbs eventologi\-cal distributions of the form
$$
\frac{p(X)}{p(\emptyset)} = \exp \Big\{ - \beta ({\mathscr
V}(X)-{\mathscr V}(\emptyset))\Big\} \frac{p_*(X)}{p_*(\emptyset)}, \ \
\ X \subseteq \frak{X},%
\eqno{(G)}
$$
$$
\frac{p(X)}{p(\emptyset)} = \exp \Big\{ \gamma ({\mathscr
V}(X)-{\mathscr V}(\emptyset))\Big\} \frac{p_*(X)}{p_*(\emptyset)}, \ \
\ X \subseteq \frak{X}.%
\eqno{(-G)}
$$
Relative entropy ``measures'' the deviation of one eventological distribution from another and reaches a minimum equal to zero when these eventological distributions coincide.

Therefore, the eventological $H$-theorem actually asserts that, of among all eventological distributions
which ``keep'' at a given level the mean value of the set-function of value $\mathscr{V}$, the Gibbs and anti-Gibbs eventological distribu\-tions lie ``closest'' to the fixed eventological distribution $p^*$.

Moreover, for the Gibbs and anti-Gibbs eventological distributions that minimize the relative entropy,
the mean value $\langle \mathscr{V} \rangle$ of set-function $\mathscr{V}$ is closely related to the relative entropy of these eventological distributions relative to $p^*$. Indeed, from ($G$) and ($-G$) it follows that
$$
{\mathscr V}(X) = -\frac{1}{\beta} \ln
\frac{p(X)}{p_*(X)}-\frac{1}{\beta} \ln
\frac{p_*(\emptyset)}{p(\emptyset)}+{\mathscr V}(\emptyset),
$$
$$
{\mathscr V}(X) = \frac{1}{\gamma} \ln
\frac{p(X)}{p_*(X)}+\frac{1}{\gamma} \ln
\frac{p_*(\emptyset)}{p(\emptyset)}+{\mathscr V}(\emptyset),
$$
Therefore
$$
\langle \mathscr{V} \rangle = \frac{1}{\beta}
\mathcal{H}_{\frac{p}{p^*}}-\frac{1}{\beta} \ln
\frac{p_*(\emptyset)}{p(\emptyset)}+{\mathscr V}(\emptyset),
$$
$$
\langle \mathscr{V} \rangle = -\frac{1}{\gamma}
\mathcal{H}_{\frac{p}{p^*}}+\frac{1}{\gamma} \ln
\frac{p_*(\emptyset)}{p(\emptyset)}+{\mathscr V}(\emptyset).
$$

\subsection{Interpretation in the language\protect\\ of conditional eventological distributions}

If we interpret:
\begin{itemize}
\item the set-function of value $\mathscr{V}$ as a characteristics of current ``market'' conjuncture, i.e. ``market'' medium that surrounds the rational subject in the ``market of perception and activity'';
\item the fixed eventological distribution $p^*$ as the past ``market'' experience of the rational subject;
\item the sought eventological distribution $p$ as a result of the interaction of the past `` market '' experience of the rational subject
with the current ``market'' environment,
\end{itemize}
then Gibbs factor and/or anti-factor
$$
\exp \left\{ -\beta \mathscr{V}(X) \right\}; \ \ \ \exp \left\{ \gamma
\mathscr{V}(X) \right\}
$$
should be interpreted as a conditional eventological distribution of the curr\-ent behavior of a rational subject under the condition of her/his past ``market'' experience and the current ``market'' conjuncture.

\textbf{Gibbs factor and anti-factor as conditional eventological} \textbf{distri\-butions}.
The formulas for the Gibbs and anti-Gibbs eventological distribu\-tions containing the Gibbs factor and anti-factor
can be looked at as the formulas of conditional probability:
$$
p(X) = p(X|Y)p^*(Y).
$$
For example,
$$
p^\downarrow(X) = \exp \left\{ -\beta(X,Y) \mathscr{V}^\downarrow(X)
\right\} p^{*\downarrow}(Y),
$$
$$
p^\uparrow(X) = \exp \left\{ \gamma(X,Y) \mathscr{V}^\uparrow(X)
\right\} p^{*\uparrow}(Y).
$$
With such an interpretation in the eventological $H$-theorem, we are talking about the fact that the conditional eventological distributions
that minimize the relative entropy have the form of the \emph{Gibbs factor}
$$
p^\downarrow(X|Y) = \exp \left\{ -\beta(X,Y) \mathscr{V}^\downarrow(X)
\right\}
$$
for events of perception and the form of the \emph{Gibbs anti-factor}
$$
p^\uparrow(X|Y) = \exp \left\{ \gamma(X,Y) \mathscr{V}^\uparrow(X)
\right\} p^{*\uparrow}(Y)
$$
for events of activity.

\subsection{Simple example}

We consider an example of functions and eventological distributions, that are present in the eventological $H$-theorem, for the monoplet of the events $\mathfrak{X} = \{x\}$:
\begin{itemize}
\item for events of perception:
$$
\mathscr{V}^\downarrow(\emptyset), \mathscr{V}^\downarrow(x);
$$
$$
\{p^{*\downarrow}(\emptyset), \ p^{*\downarrow}(x)\} =
\{1-p^{*\downarrow}(x), \ p^{*\downarrow}(x)\};
$$
$$
\langle \mathscr{V^\downarrow} \rangle =
\mathscr{V}^\downarrow(\emptyset)p^\downarrow(\emptyset)
+\mathscr{V}^\downarrow(x)p^\downarrow(x),
$$
$$
\frac{p^\downarrow(x)}{p^\downarrow(\emptyset)} = \exp \left\{ -\beta
(\mathscr{V}^\downarrow(x)-\mathscr{V}^\downarrow(\emptyset))\right\}
\frac{p^{*\downarrow}(x)}{p^{*\downarrow}(\emptyset)};
$$

\item for events of activity:
$$
\mathscr{V}^\uparrow(\emptyset), \mathscr{V}^\uparrow(x);
$$
$$
\{p^{*\uparrow}(\emptyset), \ p^{*\uparrow}(x)\} =
\{1-p^{*\uparrow}(x), \ p^{*\uparrow}(x)\};
$$
$$
\langle \mathscr{V^\uparrow} \rangle =
\mathscr{V}^\uparrow(\emptyset)p^\uparrow(\emptyset)
+\mathscr{V}^\uparrow(x)p^\uparrow(x),
$$
$$
\frac{p^\uparrow(x)}{p^\uparrow(\emptyset)} = \exp \left\{ \gamma
(\mathscr{V}^\uparrow(x)-\mathscr{V}^\uparrow(\emptyset))\right\}
\frac{p^{*\uparrow}(x)}{p^{*\uparrow}(\emptyset)}.
$$
\end{itemize}

\refen


\begin{thebibliography}{9}
\parskip=-2pt
\bibitem{6-Boltzmann1872}
{\sc Boltzmann L.} (1872) Weitere Studien \"{u}ber das
W\"{a}rmegleichgewicht unter Gasmolek\"{u}len. \emph{Wiener
Berichte}, 66: 275-370.
\bibitem{6-Vorob'ov2007}
{\sc Vorobyev O.Yu.} (2007) {\em Eventology}. --- Krasnoyarsk: SFU, 435~p (in Russian) {\footnotesize\url{https://www.academia.edu/179393/}}.
\bibitem{6-Shannon1948}
{\sc Shannon C.E.} (1949) A Mathematical Theory of Communication.
\emph{Bell System Technical Journal}, vol. 27, 379–423,
623–656.
\bibitem{6-Isihara1971}
{\sc Isihara A.} (1971) \emph{Statistical physics}. --- New York,
London: Academic Press.
\bibitem{6-Vorob'ovA2002}
{\sc Vorobyev A.O.} (2002) Multicovariances and manypoint-dependent
distributions of random sets.
{\em Proceedings of the I All-Russian FAM'2002 Conference (Oleg Vorobyev ed.). Part I}.
--- Krasnoyarsk: SFU, 21--24 (in Russian).
\bibitem{6-Vorob'ov2008}
{\sc Vorobyev O.Yu.}  (2008) Multicovariances of events.
{\em Proceedings of the VII All-Russian FAM'2008 Conference (Oleg Vorobyev ed.). Part I}.
--- Krasnoyarsk: SFU, 70--84 (in this book in Russian).
\end{thebibliography}
\end{document}